\documentclass[sigconf,screen]{acmart}
\setcopyright{none}
\usepackage{caption}
\usepackage{colortbl}
\usepackage{tikz}
\usepackage{graphicx}
\usepackage{filecontents}
\usepackage{url}

\usepackage{pifont}

\usepackage{amsmath,amssymb}
\usepackage{multirow}
\usepackage{utfsym}
\usepackage{hyperref}
\usepackage{float}                  
\usepackage{subfig}    

\usepackage{hyperxmp}
\usepackage{booktabs}
\usepackage{enumitem}
\usepackage{soul}
\usepackage{multirow}
\usepackage{booktabs}
\definecolor{lightblue}{RGB}{245, 248, 254}
\definecolor{darkblue}{RGB}{134, 134, 215}
\usepackage{cleveref}
\crefname{section}{§}{§§}
\Crefname{section}{§}{§§}

\AtBeginDocument{%
  }

\copyrightyear{2026}
\acmYear{2026}
\setcopyright{rightsretained}
\acmConference[ICSE '26]{2026 IEEE/ACM 48th International Conference on Software Engineering}{April 12--18, 2026}{Rio de Janeiro, Brazil}
\acmBooktitle{2026 IEEE/ACM 48th International Conference on Software Engineering (ICSE '26), April 12--18, 2026, Rio de Janeiro, Brazil}
\acmPrice{}
\acmDOI{10.1145/3744916.3764523}
\acmISBN{979-8-4007-2025-3/26/04}

\begin{document}
%
\title{InferLog: Accelerating LLM Inference for Online Log Parsing via ICL-oriented Prefix Caching}

\author{Yilun Wang}
\affiliation{\institution{Sun Yat-sen University}\country{Guangzhou, China}
}
\email{wangylun6@mail2.sysu.edu.cn}

\author{Pengfei Chen$^{*}$}
\affiliation{\institution{Sun Yat-sen University}\institution{Guangdong Key Laboratory of Big Data Analysis and Processing}\country{Guangzhou, China}
}
\thanks{$^{*}$Pengfei Chen is the corresponding author.}
\email{chenpf7@mail.sysu.edu.cn}

\author{Haiyu Huang, Zilong He, Gou Tan}
\affiliation{\institution{Sun Yat-sen University}\country{Guangzhou, China}
}
\email{{huanghy95,hezlong,tang29}@mail2.sysu.edu.cn}

\author{Chuanfu Zhang}
\affiliation{\institution{Sun Yat-sen University}\country{Guangzhou, China}
}
\email{zhangchf9@mail.sysu.edu.cn}
\author{Jingkai He}
\affiliation{\institution{Sun Yat-sen University}\country{Guangzhou, China}
}
\email{hejk25@mail2.sysu.edu.cn}
\author{Zibin Zheng}
\affiliation{\institution{Sun Yat-sen University}\country{Zhuhai, China}
}
\email{zhzibin@mail.sysu.edu.cn}



\begin{abstract}
Modern software systems generate massive volumes of runtime logs, necessitating efficient and accurate log parsing to enable critical downstream tasks such as anomaly detection and root cause analysis. Recently, large language models (LLMs) have achieved advanced accuracy on log parsing, but their deployment in production environments faces two major limitations: First, the privacy risks associated with commercial LLMs, driving the adoption of local deployment. Second, online log parsing poses stringent challenges for latency and throughput. While recent methods reduce the number of LLM queries, they overlook the inherent overhead of LLM inference where concurrent log parsing requests can lead to performance degradation in the LLM inference system. 

In this study, we present \textit{InferLog}, the first LLM inference optimization method for online log parsing. Our key insight is that the inference efficiency emerges as the vital bottleneck in LLM-based online log parsing, rather than parsing accuracy. \textit{InferLog} accelerates inference by designing (i) A prefix-aware ICL refinement strategy to refine the examples and permutation of in-context learning to improve the prefix caching efficiency. (ii) A rapid and task-specific configuration tuning pipeline based on meta-learning to find the optimal LLM inference system configuration. The experimental results based on Loghub-2k dataset and vLLM demonstrate that
\textit{InferLog} significantly outperforms existing inference optimization methods and markedly accelerates the state-of-the-art LLM-based log parsers without compromising parsing accuracy.



\end{abstract}

\maketitle
\section{Introduction}
\begin{figure}[t]
  \centering
  \includegraphics[width=1\linewidth]{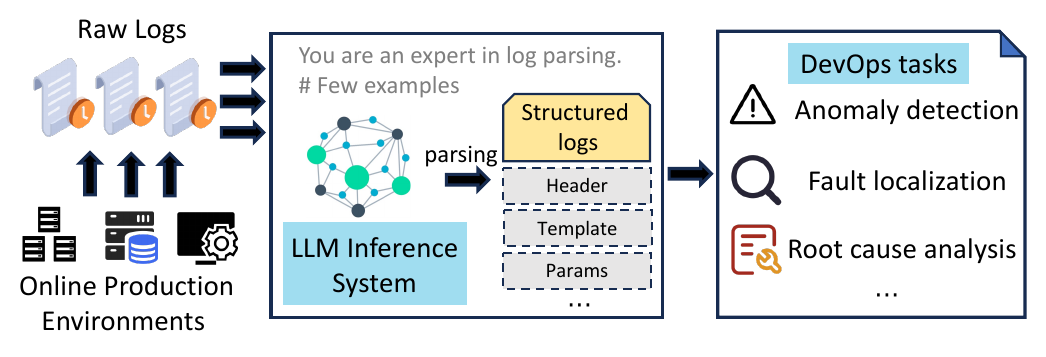}
  \caption{LLM-based online log parsing system.}
  \label{fig:pic1}
\end{figure}

Modern software systems generate massive volumes of runtime logs to record system states, events, and anomalies. Automatically parsing these semi-structured logs into structured templates, a process known as log parsing, is a foundational task in software engineering. Specifically, this task requires separating log messages into two distinct components: \textit{static templates} - constant patterns explicitly defined in logging statement and \textit{dynamic variables} - runtime-specific values reflecting system states. It enables critical downstream applications such as anomaly detection, fault localization and root cause analysis, as shown in Fig.\ref{fig:pic1}.


In recent years, large language models (LLMs) have revolutionized log parsing by leveraging their semantic understanding and few-shot learning capabilities\cite{xu2024divlog,jiang2024lilac,adparser,zhong2024logparser,llmparser-finetune,openlogparser}. Current LLM-based parsers achieve state-of-the-art (SOTA) accuracy in template extraction, even for unseen log formats. This advancement stands in contrast to traditional parsing methods, which typically rely on manually designed heuristics and syntactic analysis\cite{he2017drain,du2016spell,ael,MoLFI,yu2023brain,logppt,lenma}. Nevertheless, a critical gap persists between existing methods and modern software system deployment: 
First, commercial LLMs pose significant privacy risks for log parsing due to potential unauthorized data retention and exposure vulnerabilities\cite{openlogparser}. Consequently, organizations increasingly adopt locally deployable open-source LLMs to ensure confidential log data remains secure under strict compliance standards.
Second, a modern software system can produce log data at rates of several GB per second, imposing stringent requirements on processing \textit{latency} and \textit{throughput}\cite{logsurvey,wang2022spine}. Despite recent works have achieved promising parsing accuracy, the inference performance is far from the Service Level Objectives (SLOs) in production environments. 
In our experiments using the vLLM system, we observed that the p95 inference latency for concurrent log parsing requests was in the tens of seconds, it leads to cascading delays in downstream tasks, ultimately compromising the reliability of large-scale software systems. So, it is crucial to develop an efficient local LLM inference system to handle the massive log parsing requests and meet the SLOs.



\setlength{\abovecaptionskip}{-0.4cm} 
\setlength{\belowcaptionskip}{-0.4cm} 






Recent works\cite{jin2024ragcache,vllmapc,zheng2024batchllm} have highlighted the use of \textit{prefix caching} (PC) to enhance the performance of LLM inference systems. By sharing and reusing intermediate results (i.e., KV cache) across requests, it effectively accelerates time-to-first-token (TTFT) and reduces computation costs. While PC benefits many LLM workloads\cite{zheng2024batchllm,zhu2025subgcache,pan2025kvflow}, the effectiveness in log parsing tasks remains unexplored.
We conduct an empirical study of LLM-based online log parsing workloads under concurrent scenarios on vLLM\cite{vllm}, three critical performance issues were identified:
(i) The prefill phase (processing input sequence before generating outputs) dominates inference latency, causing 85.4\% of delays distribution, violating production SLOs. (ii) PC is ineffective in bypassing computations for tokens in in-context learning (ICL), which make up a significant portion of the context in the prompt. 
(iii) LLM system parameters, such as maximum tokens and batches, significantly impact inference performance, through automated configuration tuning, it can lead to substantial reductions in latency compared with default setting.


Generally, There are mainly two challenges in designing the inference optimization techniques for log parsing workloads.  \textbf{\textit{C1: Prefix caching imposes strict requirements on same prefix tokens}}, as the KV cache contains unique positional embeddings within the original statement and the dependencies between the tokens of the sentence. However, log parsing prompts are dynamic across requests, characterized by varying in-context learning (ICL) examples and permutation\cite{xu2024divlog}, which often lead to prefix cache missing. 
\textbf{\textit{C2: Rapid configuration adaptation for dynamic log parsing workloads}}. Logs generated by different software systems exhibit distinct characteristics and token distributions, necessitating tailored configurations for optimal performance. Existing tuning methods such as Bayesian Optimization\cite{scoot,zhang2021restune,cherrypick} and Reinforcement Learning\cite{cdbtune,li2019qtune,deepcat} demand extensive online trials. This renders them impractical for online inference tasks that demand swift configuration tuning\cite{zhu2025rockhopper}.



\textbf{\textit{InferLog} Approach}. Our key insight is that the inference efficiency emerges as the vital bottleneck in LLM-based online log parsing, which was ignored in previous works\cite{xu2024divlog,jiang2024lilac,lunar,llmparser-finetune,xiao2024free,adparser}. To this end, we propose \textit{InferLog}, a novel framework designed to optimize the LLM \underline{\textbf{\textit{Infer}}}\textit{ence} performance for \underline{\textbf{\textit{Log}}} parsing tasks, focusing on reducing latency and improving throughput in online environments. \textit{InferLog} proposes \textit{Prefix-aware ICL Refinement}(PAIR) policy to refine the log parsing prompt by adjusting the contents and order of ICL examples. Specifically, for each current request, \textit{InferLog} first identifies the examples of historical requests with the highest prefix cache hit rate probability, then performs modifying and reordering operations to update the current examples to align with it. By doing so, the ICL component can significantly boost the prefix cache hit rate(\textbf{solution to C1}).
To achieve a fast and tailored configuration optimization, we introduce \textit{Attention mechinism}\cite{attention} in \textit{Model-Agnostic Meta-Learning }(MAML)\cite{maml} to identify optimal initial parameters of the model for the target tuning tasks and then adopt \textit{Sequential Model-based Optimization} (SMBO)\cite{smbometa} to update the meta-model through few-shot learning with newly collected data to rapidly recommend high-performance configuration in a few iterations(\textbf{solution to C2}).


Generally, we make the following contributions.
\begin{itemize}[partopsep=0pt, topsep=0pt,leftmargin=2em] 
    \item We propose \textit{InferLog}, the first LLM inference acceleration framework for online LLM-based log parsing, which boosts LLM inference by \textit{prefix caching} and \textit{configuration tuning}.
    \item We identify the inefficiency of the ICL prefix cache reuse in LLM-based log parsing. To address this issue, we refine ICL demonstrations by matching, modifying and reordering to improve the prefix caching hit rate. 
    \item We introduce a fast and efficient configuration tuning pipeline that leverages attention-based MAML and SMBO to optimize LLM inference system performance.
    \item We implement \textit{InferLog} based on vLLM. The evaluations of \textit{InferLog} on the Loghub-2k dataset, demonstrate that \textit{InferLog} significantly reduces the p95 latency by 43.02\% and increases throughput by 2.14$\times$ compared to other inference acceleration baselines and is compatible to current LLM-based log parsers.

\end{itemize}


\setlength{\abovecaptionskip}{-0.5cm} 
\setlength{\belowcaptionskip}{-0.5cm} 
\begin{figure}[t]
  \centering
  \includegraphics[width=1\linewidth]{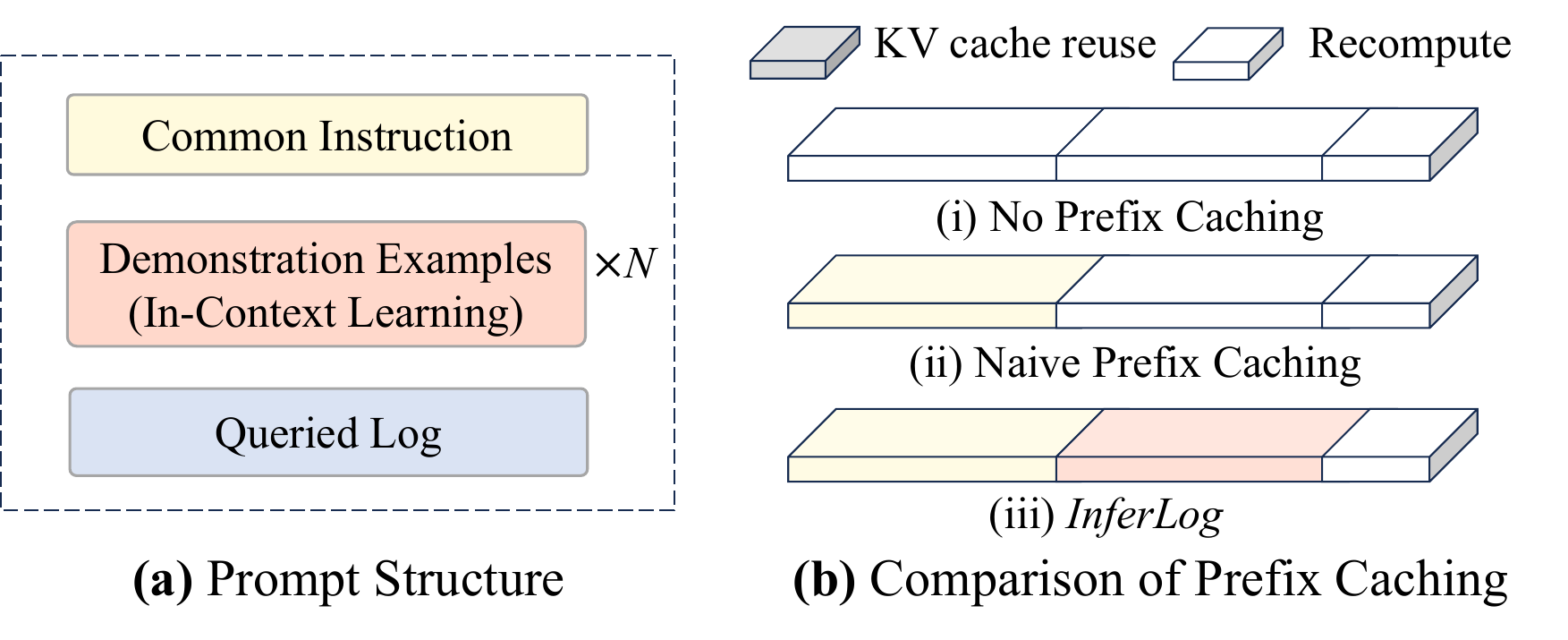}
  \caption{(a) The prompt structure of log parsing; (b) prefix caching for prompt. Colored cubes represent KV cache reuse, colorless cubes represent recompute.} 
  \label{fig:prompt}
\end{figure}

\section{Background and motivations}
\subsection{Background}
\subsubsection{Log Parsing}

Log parsing is the process of transforming semi-structured log entries into structured data by extracting both the static elements (i.e., log templates) and the variable components (i.e., log parameters) from the messages. Existing log parsers can be categorized into two main types: syntax-based and semantic-based. Syntax-based log parsers\cite{he2017drain,du2016spell,ael,MoLFI,yu2023brain} typically employ manually designed heuristics or analyze syntactic characteristics to extract log templates. However, their accuracy tends to decline when dealing with logs that do not adhere to established formats. In contrast, semantic-based log parsers\cite{liu2022uniparser,logppt,xu2024divlog,llmparser-finetune,jiang2024lilac,xiao2024free,lunar} utilize neural networks or language models to discern log templates and parameters by understanding the underlying meanings of log messages. For example, LogPPT\cite{logppt} utilizes a RoBERTa model for identifying log templates and parameters through few-shot learning. Recently, the emergence of LLMs has led to the development of various LLM-based log parsers that offer enhanced log parsing capabilities and demonstrate strong adaptability to various log formats. These LLM-based parsers employ techniques such as \textit{fine-tuning}\cite{llmparser-finetune}, \textit{prompt engineering}\cite{jiang2024lilac,xu2024divlog,adparser,zhong2024logparser} to optimize LLMs specifically for log parsing tasks, resulting in impressive performance.

\begin{figure*}[t]
\centering
\begin{minipage}[t]{0.4\textwidth}
\captionsetup{type=figure}
\centering
  \includegraphics[width=\linewidth]{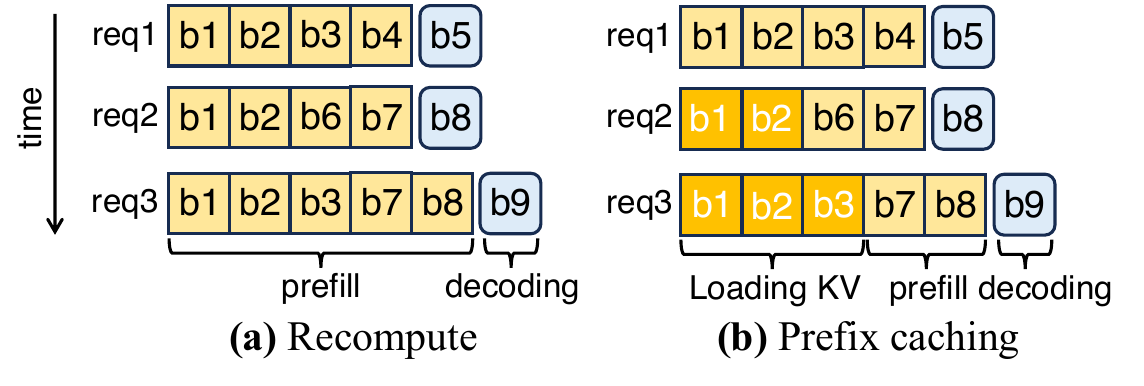}
  \caption{The comparison of recompute and PC.}
  \label{fig:prefixcache}
\end{minipage}%
\hfill%
\begin{minipage}[t]{0.28\textwidth}
\captionsetup{type=figure}
\centering
\vspace{-0.2cm}
\includegraphics[width=\linewidth]{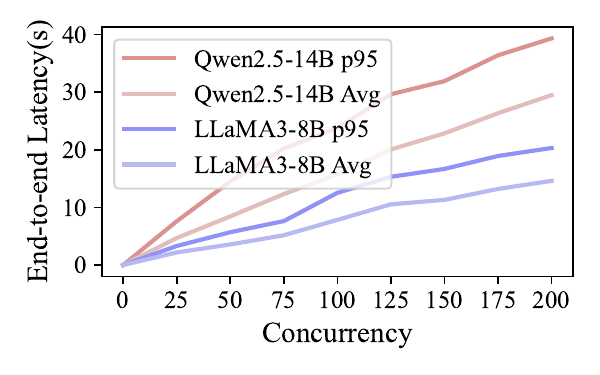}
\vspace{-0.5cm}
  \caption{Inference time under different concurrent requests.}
  \label{fig:perf}
\end{minipage}%
\hfill%
\begin{minipage}[t]{0.32\textwidth}
\captionsetup{type=figure}
\centering
\vspace{0.2cm}
  \includegraphics[width=\linewidth]{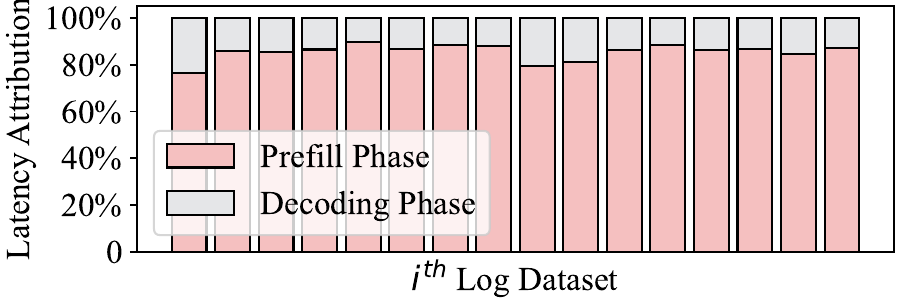}
  \caption{LLM inference latency attribution.}
  \label{fig:prefill_latency}
\end{minipage}
\end{figure*}

\textbf{ICL-based Prompt Engineering}. 
\textit{Fine-tuning} LLMs demands substantial computational resources and high-quality labeled data. Also, the high update frequency of software systems leads to frequent changes in log templates\cite{wang2022spine}, making fine-tuning less viable. \textit{In-context learning}\cite{icl-survey}, conversely, has emerged as a popular method for leveraging LLMs in downstream tasks without finetuning\cite{xu2024unilog,jiang2024lilac,xu2024divlog,zhong2024logparser,iclanomaly,sclogger}, it makes predictions based on contexts augmented with a few examples, where the selection of examples is based on the similarity to the target query. This enables LLMs to adapt quickly to new log formats and templates and reduces the dependency on training data. Every example consisted of a labeled $\{Log, Template\}$ pair. For individual log parsing, the following prompt format has been widely adopted: \textit{Common Instruction}: $\left \langle ... \right \rangle $,
\textit{Demonstration Examples}: $\left \langle ... \right \rangle $, \textit{Queried Log}: $\left \langle ... \right \rangle $, which as shown in Fig.\ref{fig:prompt}(a). 

\vspace{-0.15cm}
\subsubsection{LLM Inference System} 
The inference of LLMs is usually divided into prefill and decode phases. The prefill phases involves feeding all prompt tokens into the model in a single, parallel forward pass, generating the initial output tokens. In contrast, the decode process incrementally generates subsequent tokens based on the previously generated tokens and the input context.

\textbf{KV Cache and Prefix Caching}. 
The prefill stage is particularly time-consuming, as it requires to
compute the entire input sequence’s key (K) and value (V) tensors. Due to the autoregressive nature of model generation, each generated token depends solely on the previously generated tokens. The KV tensors computed by the attention layers are cached, and subsequent tokens use only the last generated token and the KV cache as inputs for the model's forward pass.
Prefix caching (PC)\cite{vllmapc} is an efficient technique for cross-request KV cache reuse by allowing the common KV context of prompt to be computed once and reused across different queries that share the same prefix. Specifically, if a new coming query shares the same prefix as existing one, the corresponding KV cache can be reused, allowing the new query to bypass computation for the shared tokens. Due to memory capacity, KV cache blocks are evicted based on an least recently used (LRU) policy. 

Despite these advantages, PC imposes strict requirements on same prefix tokens, underscoring the order-dependence of KV tensors.
Take Fig.\ref{fig:prefixcache} as an example, three requests arrive in order, \texttt{req2} and \texttt{req3} have some same prefix as \texttt{req1} so they can load KV cache without recomputation(i.e., [\texttt{b1,b2}] in \texttt{req2} and [\texttt{b1,b2,b3}] in \texttt{req3}), but even if \texttt{req3} and \texttt{req2} have the same logical block [\texttt{b7,b8}] in the same position, \texttt{req3} can not reuse \texttt{req2}'s KV cache of [\texttt{b7,b8}] due to missing prefix block \texttt{b6}. Without a rational management of requests can lead a poor KV cache hit\cite{zheng2024batchllm,jin2024ragcache}.



\textbf{LLM System Parameters}.
The presence of numerous configuration parameters offers flexibility for optimizing performance in inference engines\cite{vllmconf}. For example, in vLLM\cite{vllm}, \texttt{max-num-seqs} and \texttt{max-num-batched-tokens} defines the maximum number of requests and tokens the model can process simultaneously in a single inference step.
Besides, \texttt{enable-prefix-caching} and \texttt{enable-chunked-prefill} can be employed to activate respective features, which can provide significant advantages in certain scenarios.
Recent study\cite{scoot} proposes optimizing parameters for vLLM engine using Bayesian Optimization (BO), however, this approach is inefficient for online log parsing task due to the large search space and diverse token distributions. Besides, it requires much online tuning steps, and can not easily adapt to diverse workloads. \textit{InferLog} targets at the optimization of \textit{scheduling-related} parameters, which play a critical role in requests scheduling and resources utilization optimization in the environments with dynamic and continual requests.


\vspace{-0.2cm}
\subsection{Motivations}
\label{sec:motivation}
In this section, we show our empirical studies on LLM inference performance on online log parsing and introduce our motivations.
The studies in conducted on the public dataset from  Loghub-2k\cite{logpai} under vLLM\cite{vllm} inference system. Detailed information of experimental setup is provided in \cref{sec:setting}. Our study aims to answer the following research questions (RQs):
\begin{itemize}[itemsep=-1pt,partopsep=0pt, topsep=0pt,leftmargin=1em]
    \item \textbf{RQ1}: Does the LLM-based online log parsing workloads meet the requirements of SLOs under concurrent requests?
    \item \textbf{RQ2}: Does the traditional prefix caching technique achieves optimal KV cache reuse for log parsing workloads?  
    \item \textbf{RQ3}: Is the default configuration setting of LLM inference system suitable for log parsing workloads?  
\end{itemize}


\vspace{-0.1cm}
\subsubsection{LLM Inference Performance for Online Log Parsing}
\label{sec:prefill}
With the increasing complexity of software technology and distributed systems, production systems generate massive amounts of logs daily\cite{logsurvey}. Ensuring low latency to meet SLOs for online inference is crucial for downstream tasks. However, the high volume of concurrent invocations presents significant challenges to LLM inference performance. As depicted in Fig.\ref{fig:perf}, performance deteriorates significantly with an increase in concurrent requests. The average and p95 latency at a concurrency of 200 are 6.29$\times$ and 5.17$\times$ higher than at a concurrency of 25 in \texttt{Qwen2.5-14B}, and 6.70$\times$ and 6.15$\times$ in \texttt{LLaMA3-8B}, even reaching several tens of seconds. Moreover, we observe that the inference latency is predominantly attributed to the prefill phase, which accounts for an average of 85.4\% of the latency across 16 datasets in Fig.\ref{fig:prefill_latency}. 
The reason is that log parsing task involves complete instruction and ICL demonstrations in prompt, leading to the token length far exceeding the decode length, where decoding stage only outputs static log templates. So there is a pressing need to reduce prefill latency to enhance inference performance for online log parsing.
\vspace{-0.1cm}
\begin{figure}[H]
\centering
\begin{tikzpicture}
    \node[fill=lightblue, draw=darkblue, line width=0.5pt, rounded corners=2mm, minimum width=\linewidth, minimum height=1.8cm, align=center,text width=0.95\linewidth, text justified] at (0,0) {\textbf{Motivation1: }Massive concurrent invocations in LLM-based online log parsing scenario pose a substantial challenge to LLM inference performance, with the primary performance bottleneck occurring during the prefill stage.};  
\end{tikzpicture}
\end{figure}


\vspace{-0.2cm}
\subsubsection{Insufficient Prefix KV Cache Reuse}
\label{sec:moti2}
Traditional prefix caching technique, as depicted in Fig.\ref{fig:prompt}(b)(ii), optimize LLM inference by reusing KV cache for static prompt components like common instructions. However, PC can not efficiently reuse KV cache for ICL part 
due to two reasons: (i) \textit{\textbf{Dynamic examples selection}}. ICL examples are chosen based on $k$-Nearest Neighbor ($k$NN). For instance, LILAC\cite{jiang2024lilac} adopts Jaccard similarity and DivLog\cite{xu2024divlog} uses cosine distance to calculate the similarity between demonstration candidates and target log, 
leading to example sequence variations of ICL across requests. (ii) \textit{\textbf{Specific permutation}}. Existing methods\cite{jiang2024lilac,xu2024divlog} typically order the demonstrations based on their ascending similarity to the queried log, positioning the most similar demonstrations closer to it. This leads to the situation where, even different log messages select the same examples, their order within the entire demonstration
set will vary. Overall‌, these common practices in LLM-based log parsing impose prefix token sequence mismatches, rendering traditional PC ineffective for ICL tokens.
Our experiments across 16 datasets show that for naive PC, the average prefix cache hit rate\footnote{The metric of vllm:gpu\_prefix\_cache\_hit\_rate exposed by vllm.} is 55.17\% when parsing distinct log messages.

\begin{figure}[t]
  \centering
  \includegraphics[width=1\linewidth]{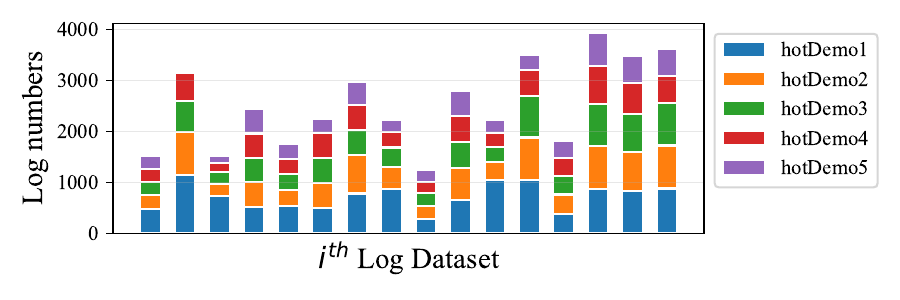}
  \caption{Selection-frequency of top 5 hotspot demonstrations for ICL chosen by 2,000 logs each dataset.}
  \label{fig:hotspot}
\end{figure}


However, although log messages are diverse, we identify the selection of ICL examples across different requests exhibits a notable degree of overlap, creating an opportunity for ICL to reuse historical KV cache. Specifically,
(i) \textit{\textbf{The candidate numbers for ICL demonstrations is limited}}. For instance, prior studies\cite{jiang2024lilac,zhong2024logparser} constrain the number of candidates to a fixed small size. (ii) \textbf{\textit{The phenomenon of hotspot demonstrations}}. We find that specific templates from the candidate set are frequently selected during the examples selection process. As illustrated in Fig. \ref{fig:hotspot}, the top five hotspot demonstration templates were selected by average 36.2\%, 27.2\%, 24.7\%, 21.2\%, and 17.1\% in each dataset\footnote{The x-axis ticks of Figure \ref{fig:prefill_latency} and Figure \ref{fig:hotspot} are the same as Figure \ref{fig:main}.}, respectively. 

\vspace{-0.1cm}
\begin{figure}[h]
\centering
\begin{tikzpicture}
    \node[fill=lightblue, draw=darkblue, line width=0.5pt, rounded corners=2mm, minimum width=\linewidth, minimum height=1.4cm, align=center,text width=0.95\linewidth, text justified] at (0,0) {\textbf{Motivation2:} Traditional prefix caching can not efficiently reuse prefix KV cache for ICL due to dynamic examples selection and specific permutation. The presence of hot demonstrations suggests that refining the ICL examples could enable it to benefit from prefix caching.};
\end{tikzpicture}
\end{figure}
\vspace{-0.3cm}
\subsubsection{Suboptimal LLM Inference Parameter Configurations Setting and Inefficient Tuning Process} 
\begin{figure}[t]
  \centering
  \includegraphics[width=1.05\linewidth]{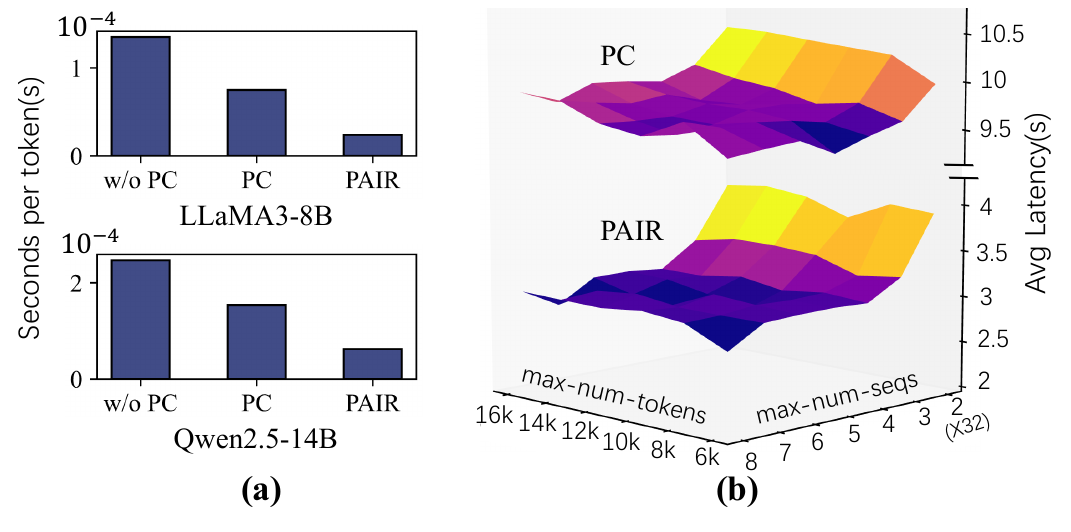}
  \caption{(a) shows the performance of different prefix caching strategies. (b) shows the performance surface of two parameters under PC and PAIR. The modification of PC strategy leads to changes in the configuration performance.}
  \label{fig:needfortune}
\end{figure}

Configuration parameters of LLM engines significantly impacts the inference performance, particularly for \textit{scheduling-related} parameters, which control the maximum batch token size and scheduling intervals, thereby determining the scheduling strategy. It is necessary to perform configuration optimization due to the following reasons. (i) \textbf{\textit{Default parameters underperform for specific workloads}}. Default parameter settings are typically designed for general use, offering a one-size-fits-all approach that may not be optimal for specific tasks. For instance, the default \texttt{max-num-batched-tokens} is often set in line with the \texttt{max-position-embeddings} of LLM, yet this setting is usually excessive for medium or short context tasks. Our experiments in \cref{sec:abla-conf} show optimized configurations can cut end-to-end latency by an average of 34.91\%, and up to 57.6\% compared to default settings.
(ii) \textit{\textbf{Impact of PC strategy on configuration governance}}. 
Under same configurations, diverse strategies demonstrate distinct performance characteristics, with PAIR cuts $\sim $6 s of latency versus PC in Fig.\ref{fig:needfortune}(b). 
By reusing KV cache in the prefill phase, it shortens per-token time, letting us pack more requests or tokens without extra latency to enhance GPU utilization.

Existing methods employ online learning techniques such as BO and RL to search configurations\cite{deepcat,cherrypick,cdbtune}. However, these approaches are time-consuming, requiring hundreds of tuning steps. This is infeasible for online inference, where the tuning time should be minimized\cite{zhang2021restune,realtimequery}. Consequently, there is a need to leverage historical knowledge to expedite the tuning process.

\setlength{\belowcaptionskip}{-0.3cm} 
\vspace{-0.3cm}
\begin{figure}[H]
\centering
\begin{tikzpicture}
    \node[fill=lightblue, draw=darkblue, line width=0.5pt, rounded corners=2mm, minimum width=\linewidth, minimum height=1cm, align=center,text width=0.95\linewidth, text justified] at (0,0) {\textbf{Motivation3: } The appropriate parameters setting is crucial for ensuring SLOs and enable an efficient LLM inference. However, existing methods are time-consuming and cannot effectively search configurations in a timely manner.};
\end{tikzpicture}
\end{figure}
\vspace{-0.3cm}
\section{METHODOLOGY}
\subsection{Overview}
\begin{figure}[t]
  \centering
  \includegraphics[width=\linewidth]{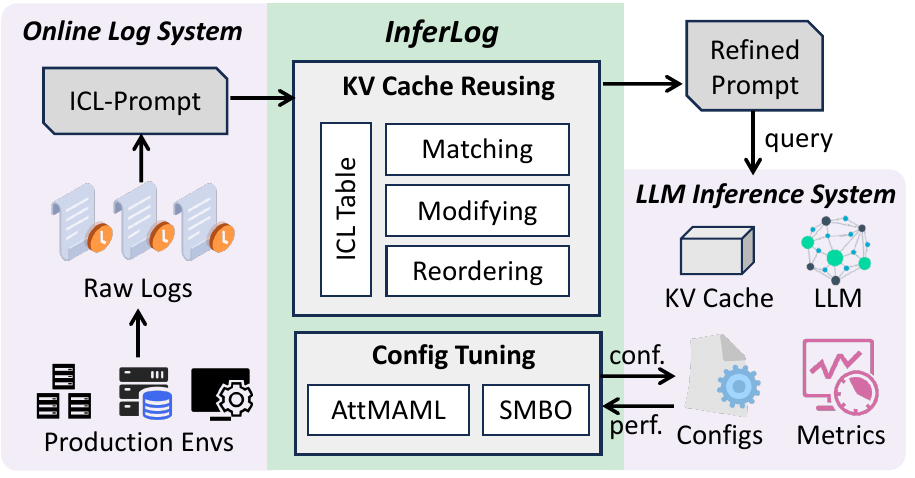}
  \caption{The architecture of \textit{InferLog}.}
  \label{fig:overview}
\end{figure}

This paper presents \textit{InferLog}, a novel framework designed to optimize the inference performance of LLM-based log parsing. The architecture, as illustrated in Fig.\ref{fig:overview}, integrates two synergistic components: an ICL-oriented KV cache reusing optimization technique to improve prefix caching efficiency (\cref{sec:novel1}), and a fast inference configuration tuning pipeline to identify optimal inference system configuration parameters to enhance performance(\cref{sec:novel2}).

Specifically, 
the software systems generate production logs, which are processed with ICL-based prompt engineering to obtain requests. Upon requests arriving, \textit{InferLog} match the historical ICL with the highest potential prefix cache hit rate from the ICL Table, then dynamically refines prompts by modifying and reordering the demonstrations of ICL to hit the prefix KV cache. 
To fast recommend optimal configurations, \textit{InferLog} proposes an attention-based meta-learning algorithm where AttMAML offline trains a good initial regression model using historical tuning data, enabling rapid adaptation to new workloads. Employing AttMAML as the surrogate model, SMBO iteratively explores the parameter space, guided by the meta-learned priors to recommend near-optimal configurations within few iterations.

\subsection{ICL-oriented KV Cache Reusing}
\label{sec:novel1}
\subsubsection{Prefix-Aware ICL Refinement}
As mentioned in \cref{sec:moti2}, there is substantial potential for ICL to benefit from prefix caching. To support this, we propose \textit{Prefix-aware ICL Refinement} 
(PAIR), which refines the demonstration examples to enhance the prefix cache hit rate of the ICL part for a more efficient LLM inference. The primary operations of the PAIR strategy include three main components: \ding{172} \textit{Prefix-based Matching}, \ding{173} \textit{Modifying} and \ding{174} \textit{Reordering}.



\begin{figure}[t]
  \centering
  \includegraphics[width=1.03\linewidth]{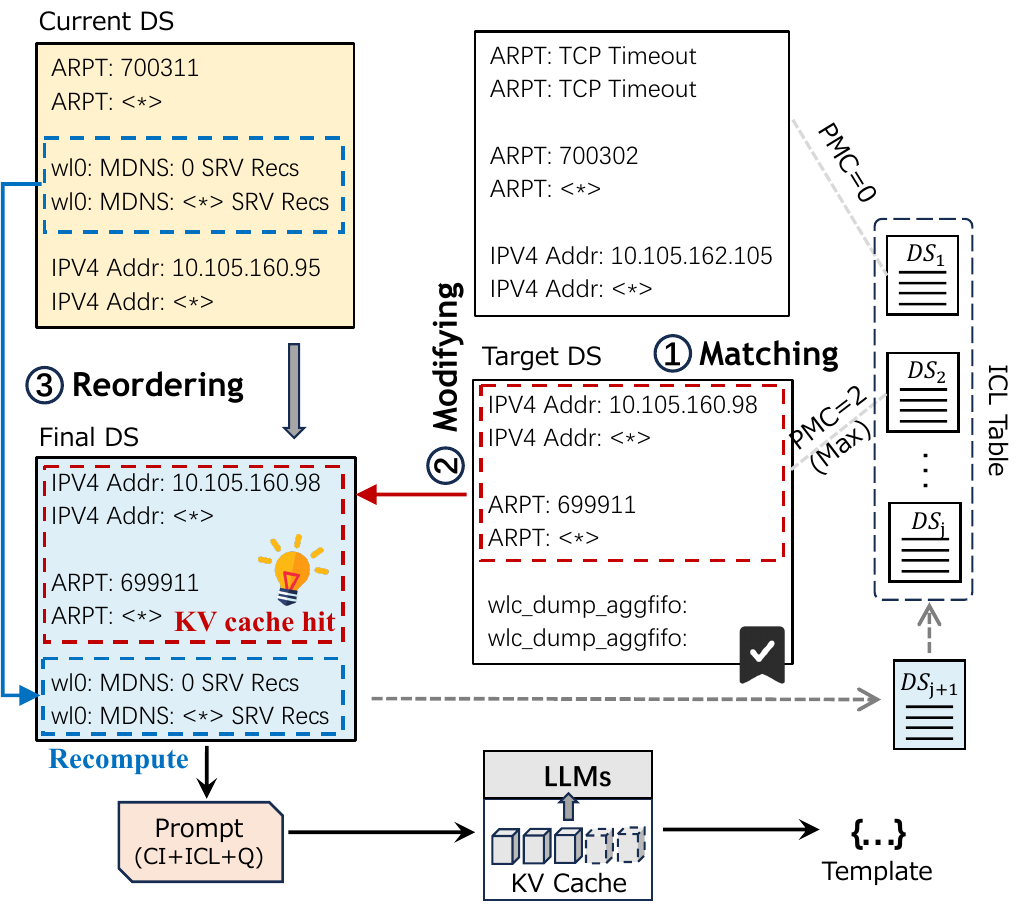}
  \caption{The demonstration cases of matching, modifying and reordering operations in PAIR.}
  \label{fig:pair}
\end{figure}

\textbf{Prefix-based Matching.}
By maintaining the \textit{ICL Table} to store historical demonstration set(DS) of ICL in each request(i.e., the request has completed inference and its KV cache have been stored in the system), which contains a demonstration set: $DS = \{D_1, D_2, ..., D_N\}$, where $D$ is an example consisting of a labeled $\{Log, Template\}$ pair. In PAIR, we first traverse the ICL Table for finding the most matching DS with maximum \textit{prefix-matching count} (PMC), as this is considered to achieve optimal KV cache reuse for the current DS.
For example, consider the scenario of Fig.\ref{fig:pair}, although both $DS_1$ and $DS_2$ share two examples with the same template with the $DS_{current}$. From the aspect of prefix, the first $D$ in $DS_1$ does not appear in $DS_{j+1}$, so the PMC for $DS_1$ is 0, indicating no possibility for prefix
cache reuse of $DS_1$. In contrast, the first two $D$ in $DS_2$ appear in current DS (i.e, \texttt{IPV4 Addr: <*>} and \texttt{ARPT: <*>}), resulting in a PMC of 2. Therefore, we mark $DS_2$ as $DS_{target}$ for prefix sharing and align current DS with $DS_{2}$. 

\textbf{Modifying}. Although $DS_{target}$ and $DS_{current}$ share similar examples with same $Template$, they do not strictly adhere to identical prefix tokens due to variations in  $Log$. Hence, it is essential to modify current DS based on $DS_{target}$ to ensure hitting prefix KV cache. For example, the current DS includes the log: \texttt{ARPT: 700311} and \texttt{IPV4 Addr: 10.105.160.95}, while $DS_2$ contains log: \texttt{ARPT: 699911} and \texttt{IPV4 Addr: 10.105.160.98}. These logs adhere to the template \texttt{ARPT: <*>} and \texttt{IPV4 Addr: <*>}. Since $DS_2$ already has the KV cache in the memory of the LLM system, we modify the current example to align with the example from $DS_{2}$ that shares the same template (indicated by the red dashed box). This operation enables examples with the same template to share the KV cache.

This operation is predicated on the assumption that demonstrations with the same template exhibit high similarity. This is because they provide semantic information that enables the LLMs to distinguish between the template and variable while remaining insensitive to specific variable values, as noted in \cite{iclwork}. This observation is further supported by the hidden representations, which yield a high cosine similarity for the representations of logs with the same template in most cases, as discussed in \cref{sec:accuracy}. 

\textbf{Reordering.}
Subsequently, we align the $DS_{current}$ with $DS_{target}$ by reordering ICL examples to achieve prefix cache reuse. As shown in Fig.\ref{fig:pair}, the original DS [$D_1, D_2, D_3$] is changed to [$D_3, D_1, D_2$]. The former portion of the final DS contains parts that can reuse KV cache, indicated by the red dashed box, whereas the latter part consists of segments that require recompute, indicated by the blue dashed box. The Final DS ensures that a significant portion of the demonstration benefit from prefix caching by directly reusing the LLM's KV cache in memory without additional computation.


\setlength{\belowcaptionskip}{-0.3cm} 

\subsubsection{ICL Table Management}

\label{sec:icltable}
Currently, \textit{InferLog} has refined the ICL to concatenate the common instruction and the queried log, forming a final prompt(Fig. \ref{fig:prompt}(a)), it then send to the LLM to perform a rapid inference to generate the log template, meanwhile, the KV cache for the cache-missing tokens is calculated and stored in memory. Due to the high cost and limited capacity of GPU memory, we can not retain the KV cache in memory permanently. 
So another important design of PAIR is the ICL Table, which serves as a bridge between ICL tokens and KV cache in LLM inference system. Specifically, we employ the \textit{OrderedDict} in Python to store the DS of historical requests due to its ability to maintain element order and support efficient reordering, and maintaining the ICL Table through a \textit{variant of LRU strategy}.

When handling a new request, we first try to perform \textit{Prefix-based Matching} to find $DS_{target}$, then update the elements in the ICL Table after processing the request. We have established the following effective rules to refine the demonstration set and update the ICL table in different cases:\\
(1) If the \textit{PMC} equals $N$ (the number of demonstrations), represents $DS_{current}$ and $DS_{target}$ use the entirely same example templates, the 
$DS_{current}$ is directly modified to $DS_{target}$ and the $DS_{target}$ is relocated to the ICL table's tail to signify recent utilization.\\
(2) If the \textit{PMC} equals 0, represents no historical DS that uses the same example template as the $DS_{current}$, the $DS_{current}$ remains unchanged due to KV cache unavailability and the $DS_{current}$ is directly appended to the ICL table's tail.\\
(3) If the \textit{PMC} fall between 1 and $N$, represents there are partially identical templates between the $DS_{target}$ and the $DS_{current}$. The refinement process of $DS_{current}$, as previously shown, will go through the methods of \textit{modifying} and \textit{reordering} to form final DS. The final DS is appended to the ICL Table's tail, while the positioning of $DS_{target}$ in ICL Table remains unchanged.


Whenever the ICL Table reaches its capacity, the eviction of the head element is carried out based on the LRU policy. Notably, \textit{InferLog} offline determines the table length based on the profiling of distribution of input tokens and remaining GPU memory which can be easily obtained in the inference system\cite{vllm}.

\subsection{Fast Inference Config Tuning}
\label{sec:novel2}
\subsubsection{Problem Formulation}
Formally, given a tuning task (log parsing workload in our scenario), our goal is to find the optimal 
LLM inference configurations ($c$) that minimize performance goals ($f$). The optimization objective is:
\vspace{-0.1cm}
\begin{equation}
c^*=arg \min_{c} f(c) ,\,subject\,\,\, to\,\,\, \sum_{k=1}^{K} f(c_k)\le T_{budget} .
\end{equation}
$f(c)$ is a black-box function, the value can be observed based on a complete workload replay. While increasing the sampling number is benefit for capture configuration-performance
plane to identify the optimal configuration, considering the requirement for near real-time responsiveness in online tuning scenario\cite{realtimequery,zhang2021restune}, the search process for sampling configurations $\left \{ c_1,c_2,...,c_k\right \} $ is under budget constraints $T_{budget}$. Although many metrics exist to evaluate performance in LLM inference systems like TTFT and TPOT\cite{scoot}, this paper focuses on optimizing total completion time for a batch of requests as a single-objective optimization process because it is easy to measure and stable to optimize.

\subsubsection{Attention MAML}
Meta-learning, also known as "learning to learn"\cite{learn2learn1,learn2learn2}, aims to acquire meta-knowledge from experiences across various tasks to mitigate the issue of requiring a large number of training samples. This paper innovatively designs the Attention MAML algorithm(AttMAML) based on the insight that tasks with similar workload features can provide the meta-learner with more valuable information\cite{zhang2021restune,dou2023turbo}.

\textbf{Model-Agnostic Meta-Learning.}
MAML\cite{maml} is a popular meta-learning framework that provides good weight initialization of a model that can achieve strong generalization to adapt to new tasks after a few gradient steps. Formally, consider a set of source tasks
$T=\left \{  T_1,T_2,...,T_M\right \} $ and initial parameters $\theta^{m}$, for each task $ T_i$ and its associated training and validation datasets $(\mathcal{D}^{i}_{train},\mathcal{D}^{i}_{val})$. In the \textit{inner-loop update} process, the task-specific parameters $\theta_{i}$ can be trained by one or more
gradient descent steps to fit $ T_i$ as follows:
\begin{equation}
 \theta_{i}\leftarrow\theta^{m}-\alpha \nabla_{\theta^{m}} \mathcal{L}^{train}_{T_{i}}\left(f_{\theta^{m}}\right),
\end{equation}
Here $\alpha$ is the base learning rate of the learner. The loss function $\mathcal{L}_{T_i}$
is defined as MSE between the predicted performance and the real one. 
In the \textit{outer-loop update}, we further calculate the expected meta-loss across all tasks according to the post-update parameters $\theta_{i}$, this meta-loss from validation data is minimized to optimize the initial parameter value $\theta^m$:
\begin{equation}
   \theta^{m} \leftarrow \theta^{m}-\beta \nabla_{\theta^{m}}  \frac{1}{M}\textstyle \sum^{M}_{i=1} \mathcal{L}^{val}_{T_{i}}\left(f_{\theta_{i}}\right), 
\end{equation}
Where $\beta$ is the meta learning rate, $M$ is the number of source tasks. MAML finds the optimal initialization weight that contains the across-task knowledge. After a few steps of fine-tune on a few-shot
dataset, the network will perform well on the unseen tasks.

\textbf{Weighted MAML}.
MAML assumes equal weights to all tasks, thus failing to discern their relative importance for the target tasks.
In real-world scenarios, a target task may be similar to only a few of the source tasks or even out-of-distribution, applying equal weighting to all sources during meta-training can be detrimental\cite{attmaml}.
To address this issue, the accuracy of meta-loss can be enhanced by employing a weighted averaging, which captures different importance between tuning tasks. The \textit{weighted meta-loss} is:
\begin{equation}
\label{eq:weightloss}
\mathcal{L}_{meta}(f_{\theta^m})=\frac{1}{W} \textstyle\sum^{M}_{i=1} w^{i}\mathcal{L}^{val}_{T_{i}}\left(f_{\theta_{i}}\right), W=\sum^{M}_{i=1} w^{i}.
\end{equation}

\textbf{Automatic Weights Learning}. Inspired by prior work\cite{gatmf}, \textit{InferLog} implements the \textit{attention mechanism}\cite{attention} to automatically learn optimal weight distributions, indicating the importance of meta-tasks in relation to target tasks. Specifically, it explores the potential impact relationships between meta-tasks and the target task by learning the dynamic attention among workload feature vectors $s$, thereby facilitating task-specific model initialization for the target domain. To acquire adaptive weights within the context of MAML, we assign a separate attention module with learnable
parameters $\theta^{att}$, the attention weight $w^i$ between target task $t$ and meta task $i$ compares the workload feature vector $s_t$ and $s_i$ through a  $\mathsf{Query}$-$\mathsf{Key}$ system:
\begin{equation}
    w^i\propto softmax(s_tW_qW^T_ks^T_i),
\end{equation}
where Query matrix $W_q$ transforms $s_t$ into a $\mathsf{query}$ and Key matrix $W_k$ transforms $s_i$ into a $\mathsf{key}$, the matching is then scaled to prevent vanishing gradients and ensure the stability of the learning process.
The parameters of the attention network $\theta^{att}$ are updated in accordance with the \textit{outer-loop update} process using the \textit{weighted meta-loss} in Eq.\ref{eq:weightloss}:
\begin{equation}
   \theta^{att} \leftarrow \theta^{att}-\beta \nabla_{\theta^{att}}   \mathcal{L}_{meta}(f_{(\theta^m,\theta^{att})}),
\end{equation}

\subsubsection{Enhancing SMBO with AttMAML}
\begin{figure}[t]
  \centering
  \includegraphics[width=1\linewidth]{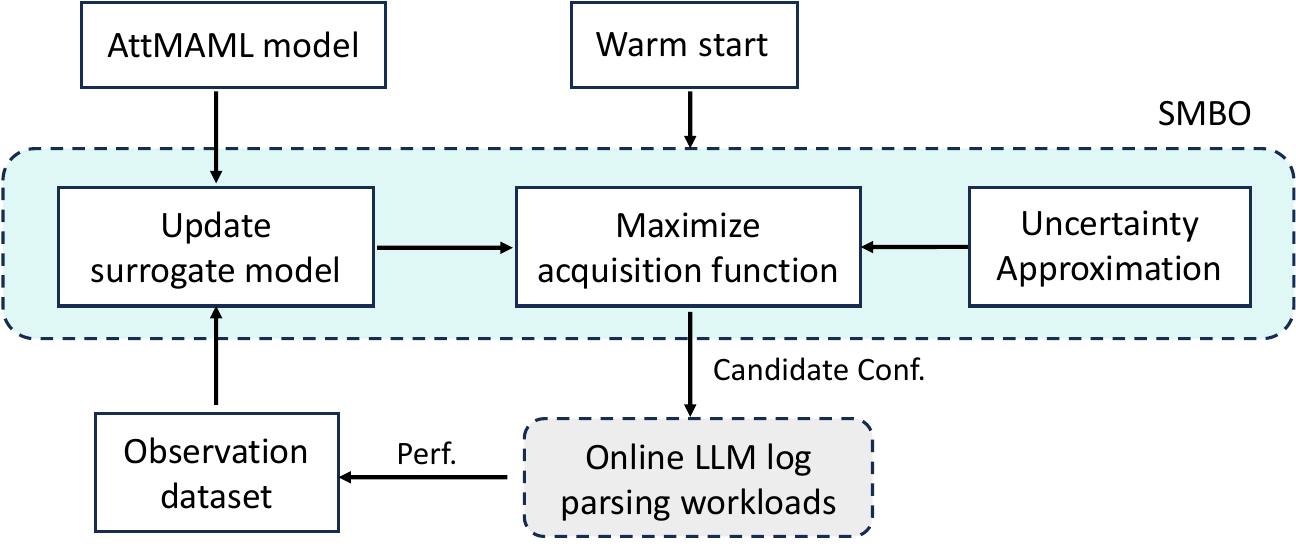}
  \caption{\textit{InferLog}'s configuration tuning workflow.}
  \label{fig:smbo}
\end{figure}

Sequential Model-based Optimization (SMBO) is a powerful method for configuration tuning\cite{smbometa,wang2021morphling}. The standard workflow of SMBO is: 1) Utilizing the \textit{surrogate model} and the \textit{acquisition function} to determine the next configuration. 2) Evaluating the newly generated configuration to obtain corresponding performance. 3) Adding the new $\left \langle configuration, performance\right \rangle$ sample into the observation dataset and then updating the \textit{surrogate model}. This process is repeated until the optimal solutions are identified or the search cost is exhausted. The workflow of configuration tuning in \textit{InferLog} is shown in Fig. \ref{fig:smbo}.

\textbf{AttMAML as Surrogate Model}.
Traditional surrogates like Gaussian Processes (GP)\cite{bogp} and Random Forest are constrained to the present task and cannot integrate insights from historical tasks, this restricts their effectiveness across diverse applications. 
\textit{InferLog} have learned a proficient initialization using AttMAML algorithm that enables the model to accurately predict the performance of configurations for new tasks after a few steps of gradient descent.
Therefore, \textit{InferLog} employs the offline trained AttMAML as the surrogate model, leveraging the tuning experience from the similar history tasks and transfering the accumulated knowledge to accelerate the tuning process of the new tasks.

Besides, following previous work\cite{dou2023turbo,fekry2020tune}, we adopt the expected improvement(EI)\cite{bishop2006pattern} acquisition function due to its effective balance between exploration and exploitation at a low computational cost.

\textbf{MC Dropout for Uncertainty Approximation}.
In classic BO, the widely used GP\cite{bogp} provides a measure of uncertainty for the predictions of unsampled data points, this uncertainty information is crucial to the optimization process as it helps balance exploration of uncertain areas with exploitation of known good regions.

To provide uncertainty analysis for SMBO, we employ the \textit{Monte Carlo Dropout} (MC Dropout)\cite{dropout4bo} strategy to obtain mean and variance information. This approach is an effective method for uncertainty quantification in neural networks.
Specifically, during the prediction phase of AttMAML model, we enable dropout, which allows us to perform multiple inferences while randomly activating different neurons each time. By calculating the mean $\mu$ and variance $\sigma^2$ of the predicted results, we can effectively assess the meta-model's uncertainty:
\begin{equation}
\mu = \frac{1}{N} \sum_{i=1}^{N} \hat{y}^{(i)},\sigma^2 = \frac{1}{N} \sum_{i=1}^{N} (\hat{y}^{(i)} - \mu)^2.
\end{equation}
where $N$ is the number of inferences performed and $\hat{y}$ presents predicted result obtained with dropout.
\setlength{\belowcaptionskip}{-0.3cm} 

\textbf{Warm Strat}.
To accelerate the convergence during the initial phase and effectively leverage successful historical knowledge, the meta-learning module proposes initial configurations by assessing task similarity, rather than relying on random sampling. In particular, we choose three samples  that represent the best configuration of top-3 most similar source tasks as initial points follow prior works\cite{sparktransfer,locat} to make the fast adaption to the new tuning task.
This not only enables the meta-model to quickly adapt to new environments using few-shot learning, but also allows for more efficient exploration guidance of the search space in SMBO process.

\section{Experimental Evaluation}
\label{sec:exp_setting}
In this section, we evaluate \textit{InferLog} to answer four questions:
\begin{itemize}[itemsep=-1pt,partopsep=0pt, topsep=0pt,leftmargin=1em]
\item \textbf{RQ1:} How does \textit{InferLog}  compare against other inference acceleration methods for online log parsing workloads? 
\item \textbf{RQ2:} How does \textit{InferLog} adapt to different LLMs and dynamic workloads?
\item \textbf{RQ3:} How does \textit{InferLog} accelerate exsiting LLM-based log parsers?
\item \textbf{RQ4:} What is the contribution of each design module?
\end{itemize}


\vspace{-0.2cm}
\subsection{Experimental Settings}
\label{sec:setting}
\textbf{Environment and Implementation.} We conduct our experiments on NVIDIA Tesla A40 48G GPUs. The CUDA version on the NVIDIA platform is 12.4. The CPU is Intel(R) Xeon(R) Gold 6326 CPU @2.90GHz, 128 GiB host memory. The operation system is Ubuntu 20.04.  We implement \textit{InferLog} with Python 3.11. We adopt vLLM-v0.6.3.post1 as the LLM inference system. We use \texttt{Qwen2.5-14B}\cite{qwen2.5} as underlying LLM because it is a relatively small yet powerful model, balancing performance and efficiency in various tasks.
In RQ2, we evaluate \textit{InferLog} by replacing \texttt{Qwen2.5} with other open-source LLMs. 

During the AttMAML training process, we divided the log datasets into two groups: half for meta-training and the remaining half for testing. To obtain prior tuning knowledge, we apply the Latin hypercube sampling(LHS)\cite{lhs} to uniformly sample 100 points within the configuration space under each meta-training task for training AttMAML. Besides, in the training phase, the sizes of the support set and query set in MAML are 15 and 45, the meta learning rate is set to 0.0001 and the base learning rate is set to 0.001. The model consists of two fully connected layers, with 64 neurons each, and the attention matrix has a dimension of 32. We report the performance of optimal configuration within 15 steps.

\textbf{Datasets.}  We assess the effectiveness of our method on log datasets from Loghub-2k\cite{logpai}. It is a widely recognized benchmark in the field of log parsing. It encompasses logs from 16 diverse systems, including distributed systems, supercomputers, operating systems, mobile platforms, server applications, and individual software packages. Each data source includes 2,000 log messages along with their respective log templates. 

\textbf{Configurations and Workloads}.
We optimize three critical \textit{scheduling-related} configuration parameters in vLLM as listed in Tab.\ref{tab:conf}, where \#\textit{Max} represents \texttt{max-position-embeddings} of LLMs. To create prompt, we samples 200 logs from each log dataset to construct the candidate set and select 5  labeled examples from the candidate set in ICL following DivLog\cite{xu2024divlog}.
We conduct optimization on a fixed batch of 2000 log parsing requests processed with a fixed concurrency level of 100 by default. We use Python $\mathsf{asyncio}$ module to simulate concurrency online requests. In RQ2, we evaluate it's adaptability under dynamic concurrency.


\setlength{\belowcaptionskip}{-0.3cm} 
\setlength{\abovecaptionskip}{0cm} 
\begin{table}[t]
\caption{Parameters to tune.}
\label{tab:conf}
\resizebox{0.8\linewidth}{!}{%
\begin{tabular}{ccc}
\toprule
\rowcolor[HTML]{dae8fc} 
\textbf{Configuration Parameter} & \textbf{Type} & \textbf{Range} \\                                               \texttt{max-num-batched-tokens}           & Integer       & {[}4000,\#\textit{Max}{]} \\
\texttt{max-num-seqs}                     & Integer       & {[}64,256{]}   \\
\texttt{scheduler-delay-factor}           & Float         & {[}0,2{]} \\ \bottomrule               
\end{tabular} 
}

\end{table}
\textbf{Metrics.}
Considering the low-latency requirements for online log parsing and the high GPU utilization of the inference system, we use the \textbf{p95 end-to-end latency} and \textbf{throughput} of requests as performance metrics for the LLM inference system. p95 end-to-end latency refers to the time within which 95\% of all requests are completed, while throughput refers to the average number of requests completed per second.
Besides, following the established metrics outlined in \cite{xu2024divlog,jiang2024lilac} for accuracy evaluation of log parsing, we employ Parsing Accuracy (PA), Precision Template Accuracy (PTA), Recall Template Accuracy (RTA), and Grouping Accuracy (GA) metrics.

\textbf{Baselines.} We compare \textit{InferLog} with the following baselines. 
(1)\textit{\textbf{Default}}: we treat vLLM\cite{vllm} with default configurations as the the basic system.
(2)\textit{\textbf{PC}}\cite{vllmapc} uses default prefix caching technique for optimizing inference latency in vLLM.
(3)\textit{\textbf{PC-Tune}}: we adopt random search\cite{rs} to optimize inference configuration parameters based on PC.
(4)\textit{\textbf{PC-Chunk}}\cite{chunk} optimizes context filling efficiency by processing input text in chunks, enhancing generation speed and memory usage.
(5)\textit{\textbf{BatchLLM}}~\cite{zheng2024batchllm} designs prefix-sharing group based scheduling and request groups reordering strategies to improve the prefix caching ratio. Beside, it also determines the best chunk size of chunked-prefills\cite{chunk} policy to achieve the optimal throughput.
Notably, except for the default, all other approaches have the prefix caching option enabled.
 
\setlength{\belowcaptionskip}{-0.3cm} 
\begin{figure*}[t]
  \centering
  \includegraphics[width=1\linewidth]{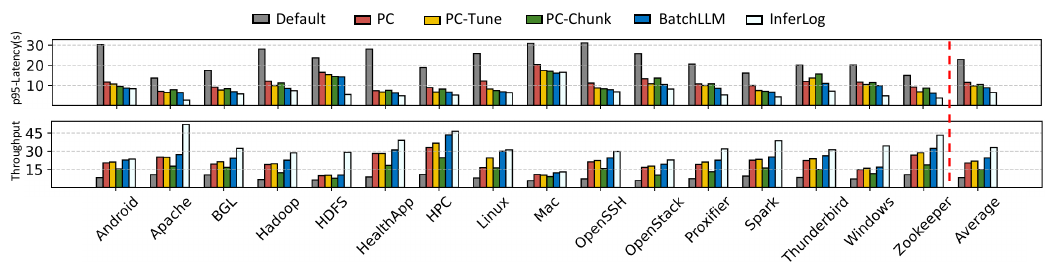}
  \caption{Performance of \textit{InferLog} compared to other inference optimization methods on public datasets (Lower is better for p95-latency, higher is better for throughput).}
  \label{fig:main}
\end{figure*}
\vspace{-0.2cm}
\section{Results and Analysis}
\subsection{Effectiveness Validation (RQ1)}
\label{sec:rq1}
\subsubsection{Inference Performance Improvement}
Fig.\ref{fig:main} provides a comparative analysis of
various inference optimization methods across 16 datasets in terms of latency and throughput. Experimental results show that \textit{InferLog} significantly outperforms other methods.  Specifically, \textit{InferLog} achieves a 71.9\%\, 43.9\%, 34.6\%, 38.0\% and 26.7\% reduction in p95 latency, as well as a 4.02$\times$, 1.62$\times$, 1.51$\times$, 2.21$\times$, 1.35$\times$ speedup on throughput compared to default vLLM\cite{vllm}, PC\cite{vllmapc}, PC-Tune, PC-Chunk\cite{chunk} and BatchLLM\cite{zheng2024batchllm} on average, respectively.

Specifically, the default inference does not employ any optimization techniques, and the system fully executes the request inference process, resulting in the worst inference performance.
Due to the fact that numerous log parsing requests with same log templates in log parsing task, they tend to opt for the similar demonstrations in ICL. Therefore, employing PC can lead to a significant acceleration in inference performance. However, PC, PC-Tune and PC-Chunk can only strictly reuse the KV cache with identical prefixes, which limits the reuse of KV cache between requests. PC-Chunk\cite{chunk} utilize chunk-prefill\cite{chunk} technique to divide requests into chunks and execute them in parallel. However, the chunking operation introduces additional overhead, which is not conducive to the execution of short requests like log parsing, so it leads to a lower throughput than PC and PC-Tune.
BatchLLM\cite{zheng2024batchllm} performs better than other methods because it reorders the requests to improve the prefix cache hit rate. However, BatchLLM do not perform fine-grained ICL reordering within individual requests, leading to suboptimal KV cache reuse. Additionally, the adjustment of request order also introduces waiting delays. In contrast, \textit{InferLog} adjusts the ICL examples of each incoming request, enabling optimal prefix cache utilization, and further optimizes the configuration to enhance performance. 

\textbf{\textit{Overall, the
experimental results demonstrate that \textit{InferLog} is more effective than existing inference optimization methods and can be applied to various online log parsing requests.}}

\subsubsection{Parsing Accuracy}
As mentioned in \cref{sec:moti2}, existing LLM-based parsers use ascending order to arrange ICL examples\cite{xu2024divlog,jiang2024lilac}, which has been proven to perform better than the descending and random orders. However, \textit{InferLog} adopts the PAIR by modifying and reordering the examples to enhance the prefix KV cache hit rate, thereby accelerating the inference process.
Tab.\ref{tab:accuracy} provides a comparative analysis of \textit{w/o Inferlog} and \textit{w/ Inferlog} across 16 datasets in terms of PA, PTA, RTA, and GA metrics. we marked the best results in each metric in bold font.
Compared to original ICL examples with ascending permutation(\textit{w/o InferLog}), \textit{w/ InferLog} achieves comparable log parsing accuracy. Specifically, the changes in the four metrics of \textit{InferLog} compared to \textit{w/o InferLog} are +0.7, -0.1, +0 and +0.3. 
\textbf{\textit{This indicates that InferLog achieves faster LLM inference without sacrificing log parsing accuracy}}.
\vspace{-0.3cm}
\subsection{Generalizability Analysis (RQ2)}
\label{sec:rq3}
\subsubsection{Performance Under Dynamic Workloads}
\begin{figure}[h]
  \centering
  \includegraphics[width=\linewidth]{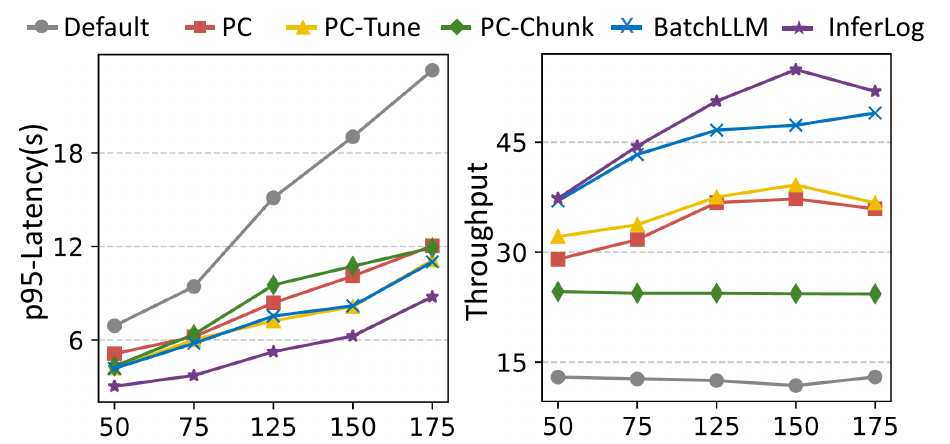}
  \caption{P95-latency and throughput on different concurrency.}
  \label{fig:workload}
\end{figure}

In real production, the workload always changes over time. We select the concurrency of 50, 75, 100, 125, 150 and 175 to evaluate it's adaptability on different workloads, the results are shown in Fig.\ref{fig:workload}. It is evident that \textit{InferLog} performs well across all concurrency settings,
compared to other methods, \textit{InferLog} reduces the p95 latency by 42.43\% and increases 1.58$\times$ throughput across all concurrencies on average.
\textit{\textbf{The results demonstrate that \textit{InferLog} can adapt well to varying workloads and achieve stable performance improvements.}}
\vspace{-0.1cm}
\subsubsection{Performance Under Different LLMs} 
\begin{figure}[h]
  \centering
  \includegraphics[width=1\linewidth]{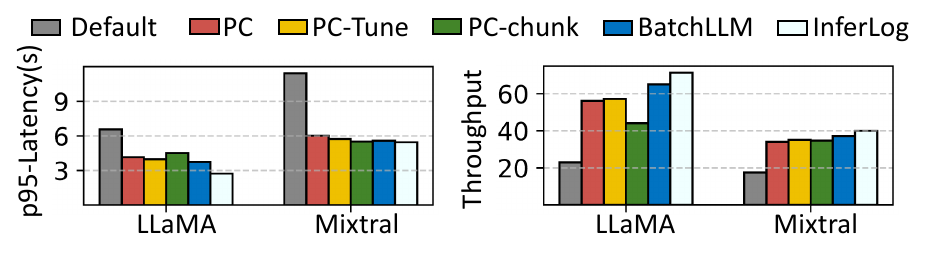}
  \caption{P95-latency and throughput on different LLMs.}
  \label{fig:model}
\end{figure}

To validate LLM generalization, we evaluate two representative LLMs with divergent architectures: (1) \texttt{LLaMA3-8B}\cite{llama3modelcard}, and (2) the sparse Mixture-of-Experts (MoE)-based \texttt{Mixtral-7B}\cite{jiang2024mixtral}. As shown in Fig.\ref{fig:model}, \textit{InferLog} achieves consistent performance enhancements across evaluated models. Specifically, it surpasses the baselines by 40.53\% reduction on p95 latency and 1.45$\times$ throughput for \texttt{LLaMA3-8B}. Also, \textit{InferLog} achieves 20.52\% reduction on p95 latency and 1.26$\times$ throughput for \texttt{Mixtral-7B}. \textit{\textbf{This provides compelling evidence that \textit{InferLog} is capable of effectively accelerating a variety of LLMs with diverse architectures. }}
\setlength{\belowcaptionskip}{-0.1cm} 
\setlength{\abovecaptionskip}{-0.1cm} 
\begin{table}[t]
\centering
\caption{Log parsing accuracy comparison.}
\label{tab:accuracy}
\resizebox{\linewidth}{!}{%
\begin{tabular}{c|cccc|cccc}
\toprule[1pt]
 & \multicolumn{4}{c|}{\textbf{w/o \textit{InferLog}}} & \multicolumn{4}{c}{\textbf{w/ \textit{InferLog}}} \\
 & PA & PTA & RTA & GA & PA & PTA & RTA & GA \\
\hline
Android     & \textbf{97.4} & 86.3   & 88.0 & \textbf{98.9}    & \textbf{97.4} & \textbf{87.0} & \textbf{90.0} & 97.4  \\
Apache      & \textbf{100.0}     & \textbf{100.0}        & \textbf{100.0}       & \textbf{100.0}         & \textbf{100.0}      & \textbf{100.0}        & \textbf{100.0}       & \textbf{100.0}      \\
BGL         & \textbf{97.8} & \textbf{94.2} & \textbf{95.8} & \textbf{99.3}     & \textbf{97.8} & 92.7   & 95.0     & 98.3 \\
Hadoop      & 92.9  & \textbf{91.2} & \textbf{91.2} & \textbf{100.0 }        & \textbf{99.3} & 88.6 & 88.6 & 99.8  \\
HDFS        & \textbf{99.8} & \textbf{75.0}     & \textbf{85.7}    & \textbf{82.9}    & 99.6  & 75.0     & \textbf{85.7}    &\textbf{82.9} \\
HealthApp   & 97.5 & 87.3 & 92.0     & 87.8    & \textbf{98.1} & \textbf{93.4 }&\textbf{ 94.7} & \textbf{99.2} \\
HPC         & 98.4  & \textbf{75.0}    & \textbf{84.8} & \textbf{94.9}     & \textbf{99.3} & 71.7 & 82.6 & 93.4 \\
Linux       & \textbf{99.8} & \textbf{95.6}   & \textbf{95.6}   & \textbf{100.0}         & 99.6 & 93.0 & 92.2  & 99.9 \\
Mac         & 72.3  &\textbf{ 64.2} & 72.1   & \textbf{79.2}    & \textbf{74.2} & 63.0  & \textbf{73.0} & 76.9 \\
OpenSSH     & 93.2 & \textbf{88.9} & \textbf{92.3} & \textbf{93.3}   & \textbf{93.3}  & \textbf{88.9}& \textbf{92.3} & \textbf{93.3} \\
OpenStack   & \textbf{99.8}  & \textbf{90.7} &\textbf{95.1} & \textbf{97.9}    & \textbf{99.8} & 86.4 & 92.7 & 96.8  \\
Proxifier   & \textbf{99.9} & \textbf{69.2}   &\textbf{75.0}     & \textbf{67.8}     & 99.7 & 64.3 & \textbf{75.0}     & 54.4  \\
Spark       & \textbf{99.9} & \textbf{94.4} & \textbf{94.4} & \textbf{100.0}         & \textbf{99.9} & \textbf{94.4} & \textbf{94.4} & \textbf{100.0}      \\
Thunderbird & \textbf{88.9}  & 83.0     & 85.2 & 98.3     & \textbf{89.9} & \textbf{86.2}& \textbf{87.9}& \textbf{99.0} \\
Windows     & 99.0 & 74.0 & 80.0     & 99.4    & \textbf{99.3}& \textbf{82.3}     & \textbf{84.0}     & \textbf{99.9} \\
Zookeeper   & \textbf{99.9 }& 90.4 & \textbf{94.0}     & 84.9 & \textbf{99.9} &\textbf{ 92.2} & \textbf{94.0}    & \textbf{99.5}\\
\hline
Average & 96.0 & \textbf{85.0} & \textbf{88.9} & 92.8 & \textbf{96.7} & 84.9 & \textbf{88.9} & \textbf{93.1}\\
\bottomrule[1pt]
\end{tabular}
}
\end{table}
\setlength{\belowcaptionskip}{-0.4cm} 
\vspace{-0.3cm}
\setlength{\belowcaptionskip}{-0.3cm} 

\subsection{Compatibility Evaluation (RQ3)} 
In this RQ, our primary focus is on assessing the compatibility of our method with current LLM-based log parsers. Specifically, recent studies have proposed methods to enhance the efficiency of LLM-based log parsing by analyzing the characteristics of log parsing tasks. For instance, LILAC\cite{jiang2024lilac} adopts an adaptive parsing cache to store the template to prevent duplicate queries of the LLM, while LogBatcher\cite{xiao2024free} employs a clustering method and generate log template for each partition. 
Therefore, we want to see if our method can work together with the current approaches and further improve their inference efficiency. We evaluate the performance of current LLM-based log parsers: DivLog\cite{xu2024divlog}, LILAC\cite{jiang2024lilac} and LogBatcher\cite{xiao2024free} with the acceleration by the \textit{InferLog}. For DivLog, we query all log parsing requests in the dataset. For LogBatcher, we first offline partition logs and perform only one query of each partition. For LILAC, we only query LLM for log parsing requests with different templates because it caches historical templates.

The average results across 16 datasets are illustrated in Tab.\ref{tab:compatible}. It is obvious that with \textit{InferLog}, the performance of current log parsers has been significantly enhanced. Specifically, for DivLog,
the performance improvement is the most pronounced, the reason is that a large batch of queries have brought performance bottlenecks to the LLM inference system.
Through PAIR and configuration optimization by \textit{InferLog}, the latency is reduced from 22.89s to 6.43s, marking a 71.90\% decrease, while the throughput improvement is 301.83\%. 
For LILAC and LogBatcher, they eliminate redundant LLM queries and focus on parsing unseen logs, the average number of log parsing queries to the LLM is reduced by $\sim$23$\times$. Nonetheless, \textit{InferLog} still achieves a performance improvement of 50.17\% and 49.94\% in latency reduction and 66.07\% and 78.31\% in throughput increase for LILAC and LogBatcher, respectively. \textit{\textbf{Overall, InferLog can improve the efficiency of LLM-based log parsers and is compatible with existing techniques that aim to reduce numbers of LLM invocation.}}




\setlength{\abovecaptionskip}{0cm} 
\setlength{\belowcaptionskip}{0cm} 
\begin{table}[h]
\centering
\caption{Results of accelerating LLM-based log parsers.}
\label{tab:compatible}
\resizebox{0.85\linewidth}{!}{
\begin{tabular}{lcc}
\toprule[1pt]
 & \textbf{p95-Latency}(s) & \textbf{Throughput} \\
\hline
DivLog & 22.89 &8.20  \\
DivLog w/ \textit{InferLog} &  6.43 ($\downarrow$ 71.90\%) & 32.95 ($\uparrow$ 301.83\%) \\
\hline
LILAC & 12.30 & 5.04 \\
LILAC w/ \textit{InferLog} & 6.13 ($\downarrow$ 50.17\%) & 8.37 ($\uparrow$ 66.07\%)  \\
\hline
LogBatcher  & 15.37 & 4.92 \\
LogBatcher w/ \textit{InferLog} & 7.69 ($\downarrow$ 49.95\%) & 8.77 ($\uparrow$ 78.31\%)\\
\bottomrule[1pt]
\end{tabular}
}
\end{table}

\vspace{-0.3cm}
\subsection{Ablation Study (RQ4)} 
This RQ gives a comprehensive explanation of each module’s contribution, i.e., the PAIR in  and the inference configuration tuning. We create the following
two variants of \textit{InferLog} and compare them with the original approach. (1) \textit{InferLog} w/o PAIR: remove
the ICL refinement and only keep default PC. (2) \textit{InferLog} w/o Config Tuning: remove configuration tuning to default setting. Tab.\ref{tab:ablation} shows the average results cross 16 datasets. It is clear that removing any of the two modules will affect performance to some extent.
\begin{table}[t]
\centering
\caption{Ablation study results.}
\label{tab:ablation}
\resizebox{0.85\linewidth}{!}{
\begin{tabular}{lcc}
\toprule[1pt]
 & \textbf{p95-Latency}(s) & \textbf{Throughput} \\
\hline
\textit{InferLog} & 6.43 & 32.95  \\
\textit{InferLog} w/o PAIR &  9.45 ($\uparrow$ 46.97\%) & 22.4 ($\downarrow$ 32.02\%) \\
\textit{InferLog} w/o Config Tuning & 9.88 ($\uparrow$ 53.65\%) & 28.16 ($\downarrow$ 14.54\%) \\
\bottomrule[1pt]
\end{tabular}
}
\end{table}
\subsubsection{Contribution of PAIR}
The dynamic nature of ICL in log parsing poses significant challenges for benefiting from PC techniques. PAIR is a crucial module that
enables a significant part of ICL tokens to leverage the historical KV cache, thereby bypassing redundant computations cross online log parsing requests. As shown in Tab.\ref{tab:ablation}, when PAIR is removed, LLM can only reuse the KV cache of common instruction in user requests. Without it, the inference latency increases by 46.97\% and processing throughput drops by 32.02\%. 

\textbf{PC Hit Rate}. To better understand the performance drop, Fig.\ref{fig:hitrate} illustrates the average PC hit rates across all datasets. Obviously, PAIR achieves the highest hit rate, with average 1.23$\times$, 1.16$\times$ and 1.18$\times$ higher than naive PC\cite{vllmapc}, PC-Chunk\cite{chunk} and BatchLLM\cite{zheng2024batchllm}, and reaching up to 2.1$\times$ on the HDFS dataset compared with PC(Omitted from figures), confirming the effectiveness of PAIR.
\setlength{\belowcaptionskip}{-0.3cm} 
\setlength{\abovecaptionskip}{-0.3cm} 
\begin{table}[htbp]
    \centering
    \begin{minipage}[t]{0.23\textwidth}
        \centering
        \includegraphics[width=1.05\textwidth]{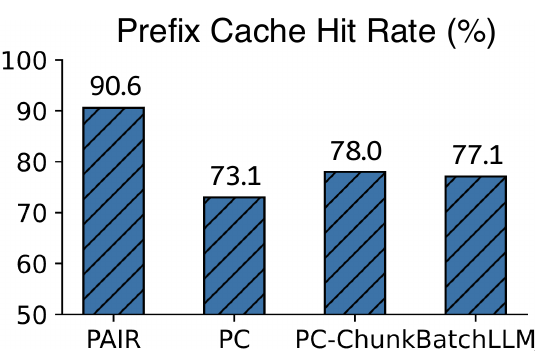}
        \captionof{figure}{Prefix cache hit rate.}
        \label{fig:hitrate}
    \end{minipage}
    \hfill
    \begin{minipage}[t]{0.23\textwidth}
        \centering
        \includegraphics[width=1.\textwidth]{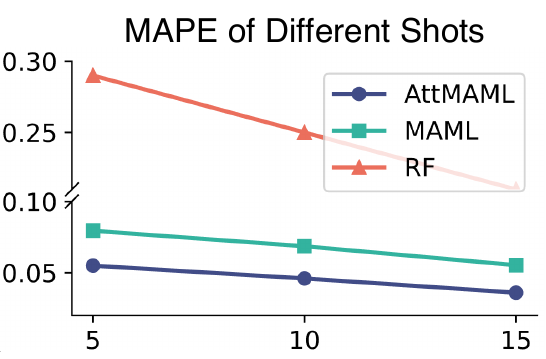}
        \captionof{figure}{MAPE of few-shot learning.}
        \label{fig:mape}
    \end{minipage}
\end{table}

\begin{figure}[h]
  \centering
  \includegraphics[width=0.6\linewidth]{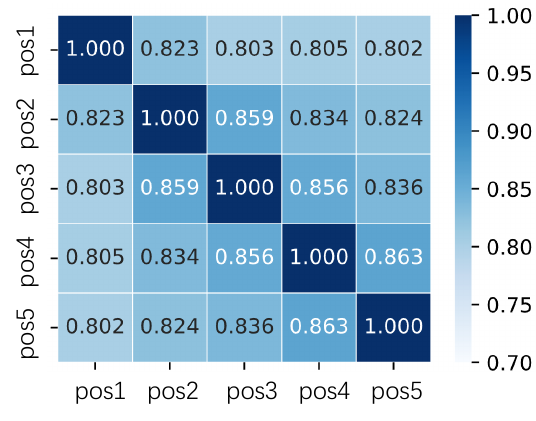}
  \caption{The heatmaps of similarities in representations of prompt from the last layer outputs across different ICL positions.}
  \label{fig:heatmap}
\end{figure}

\vspace{-0.2cm}
\textbf{Impact on Parsing Accuracy}.
\label{sec:accuracy}
To delve deeper into the parsing accuracy, we analysis the hidden representations when implementing PAIR by \texttt{Qwen2.5-14B}.
First, we compared the representations of input tokens where examples were positioned in various orders, which reflects the impact of \textit{Reordering}. To create different orders, we adopt the same setting
from prior work\cite{iclorder}. As depicted in Fig.\ref{fig:heatmap}, it presents heatmaps of average representation similarities across different example positions of the Apache dataset. It is noteworthy that the cosine similarity across representations at any position typically exceeded 0.80 and up to 0.86, that is considered to possess sufficient robustness to counterbalance the demonstration order sensitivity of decoder-only LLMs according to prior work\cite{iclorder}. 
We believe that with the enhancement of LLMs capability and carefully designed prompt engineering, LLMs' context understanding and noise resistance will improve, rendering them more robust to variations in ICL order.  
Second, we compared ICL after \textit{Modifying} with the original ICL on the entire prompt representation. The average similarity of output representations from the last layer is 0.9977 and 0.9983 for HealthApp and Apache dataset, respectively (Omitted from figures). In general, the semantic impact of PAIR on the prompt is not obvious, and it still maintains the high level log parsing accuracy. 

\vspace{-0.1cm}
\setlength{\belowcaptionskip}{-0.3cm} 
\setlength{\abovecaptionskip}{-0.3cm} 

\setlength{\belowcaptionskip}{-0.5cm} 
\setlength{\abovecaptionskip}{-0.5cm} 
\begin{figure}[t]
  \centering
  \hspace{-0.2cm}
  \includegraphics[width=1\linewidth]{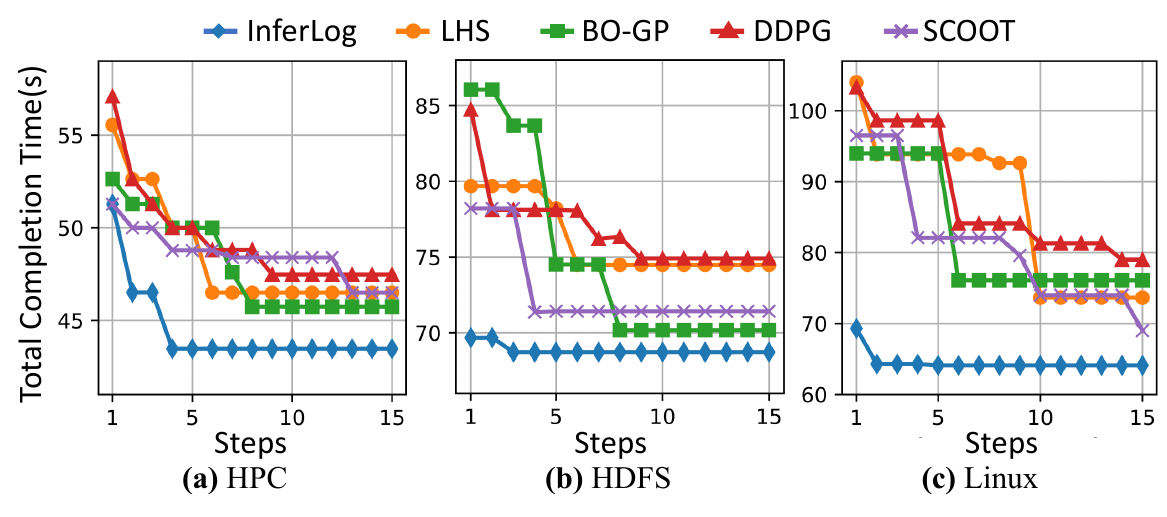}
  \caption{The current best performance along with the 15
online tuning steps (lower is better).}
  \label{fig:tune}
\end{figure}
\vspace{-0.1cm}
\subsubsection{Contribution of Configuration Tuning}
\label{sec:abla-conf}
Inference configurations directly influence the request scheduling and execution performance of LLMs, \textit{InferLog} aims to provide an efficient and highly transferable configuration tuning pipeline to further enhance the performance of inference systems.
Without configuration tuning, the inference latency degrades by 53.65\% and processing throughput drops by 14.5\%. Notably, a good configuration achieves significant improvements in tail latency, which is crucial for meeting SLOs. Indeed, it limits the tokens processed per inference step, thereby preventing computational overload in prefill phase, 
which effectively alleviates the performance bottlenecks identified in \cref{sec:prefill}.

\textbf{Prediction Accuracy of AttMAML}.
\textit{InferLog} innovatively integrates the attention mechanism into meta-learning framework, enabling the MAML algorithm to automatically identify which meta-training tasks are more critical for the target tasks. This enhancement not only boosts the precision of the MAML algorithm but also enhances its ability to rapidly adapt to target tasks. Specifically, we conducted meta-learning on eight tasks and evaluated the few-shot learning performance on another eight tasks. The average mean absolute percentage error (MAPE) on target tasks are shown in Fig.\ref{fig:mape}, we find that (i) compared to Random Forest(RF), the representation of machine learning, AttMAML demonstrates a 81.76\% reduction on MAPE, indicating it's ability to transfer knowledge effectively in facing new scenarios. (ii) AttMAML achieves a 32.74\% MAPE reduction over the standard MAML algorithm on average, emphasizing the effectiveness of the attention mechanism in improving meta-training performance.

\textbf{Online Tuning Efficiency}.
Efficiency is a crucial consideration in recent configuration tuning methods\cite{deepcat,shen2023rover,locat}. To evaluate it, we compare \textit{InferLog} with several prominent algorithms: LHS\cite{lhs}, BO with Gaussian Processes (BO-GP)\cite{bogp}, DDPG\cite{ddpg}, and the SOTA method for LLM configuration tuning SCOOT\cite{scoot}, the optimization was conducted within a sampling budget of 15 steps. We show tuning results of three datasets in the Fig.\ref{fig:tune}. 
\textit{InferLog} outperforms other methods by requiring 40.74\%, 53.85\% and 79.68\% fewer iterations to determine the optimal configuration on HPC, HDFS and Linux, resulting in a performance improvement of 1.07$\times$, 1.05$\times$ and 1.13$\times$.
The reason is that \textit{InferLog} uses a warm-start strategy to leverage historical optimal configurations from similar workloads to achieve rapid early convergence.
Then, AttMAML-enhanced SMBO enables the meta-model to be fine-tuned in just a few steps within new environments, efficiently guiding the configuration tuning process.

\section{Discussion}
\subsection{Practicality and Specialization}
\textbf{\textit{InferLog} is orthogonal with LLM-based log parsers}.
Recent LLM-based log parsers\cite{xu2024divlog,jiang2024lilac,lunar,adparser,xiao2024free,zhong2024logparser} design log parsing algorithms such as in-context learning, template caching or clustering, aiming to improve parsing accuracy. Conversely, \textit{InferLog} targets the underlying LLM inference system, it acts as an intermediate layer between log parsers and the LLM inference system, \textit{InferLog} is agnostic to log parsers and seamlessly integrates with them to improve inference efficiency via prefix KV cache reusing and configuration tuning without compromising accuracy. \\
\textbf{\textit{InferLog} achieves higher efficiency in log parsing than in other LLM inference workloads}.
Although methods like KV cache reusing and configuration tuning in \textit{InferLog} can accelerate LLM inference in other contexts, there are two main features in log parsing make \textit{InferLog} specifically suitable for log parsing and may lead to performance degradation in other scenarios.
Firstly, the order of ICL examples in log parsing has little impact on \textit{InferLog}'s results. This is mainly because log parsing exhibits greater predictability and stability in its output structure, focusing on extracting log templates, so the model's output is less affected by the order of examples. In contrast, in RAG or document-based inference, the relative index of information within the document is more critical for accurate answers\cite{jin2024ragcache,liu2023lost}. Secondly, configuration tuning process requires a stable mapping between configuration and performance metrics. Log parsing task outputs stable log templates with predictable token length, creating a foundation for configuration tuning. However, in general workloads, the output lengths of LLMs are dynamic(e.g., varying with input context or task requirements), disrupting tuning model convergence.


\vspace{-0.1cm}
\subsection{Threats to Validity}
We have identified the following threats to validity:

\textbf{Subject Systems}. We conduct experiments on the vLLM inference system, due to its high popularity. We  also compare the inference performance across various open-source LLMs. \textit{InferLog} only requires the target inference system and LLMs to support prefix caching, without the need for system modifications. Overall, \textit{InferLog} can efficiently perform log parsing tasks on locally deployed systems and open-source LLMs without privacy concerns.

\textbf{Randomness}. Randomness may affect the performance through two aspects: the randomness of LLMs output and the
randomness introduced during the performance testing of the system, particularly under online concurrent requests where task scheduling introduces variability in system performance.
To mitigate the former threat, we set the \texttt{temperature} to 0, guaranteeing consistent outputs for identical input text. 
To mitigate the latter threat, each experiment was repeated three times for every experimental setting, and we report the the average results as the final outcome.
\vspace{-0.3cm}
\section{Related Work}


\textbf{LLM-based Log Parsing.}
Log parsing is a critical preliminary step for various log analysis
tasks. Therefore, numerous efforts have been made to achieve accurate log parsing\cite{he2017drain,du2016spell,ael,MoLFI,yu2023brain,lenma,logsig}. With the rise LLMs, a series of LLM-based log parsers have been developed to achieve more effective log parsing\cite{xu2024divlog,jiang2024lilac,lunar,adparser,llmparser-finetune,openlogparser,zhong2024logparser,xiao2024free}. These LLM-based log parsers
leverage fine-tuning\cite{llmparser-finetune}, in-context learning\cite{jiang2024lilac,xu2024divlog,adparser}or log clustering\cite{xiao2024free,adparser} to specialize LLMs for log parsing tasks. Considering high invocation costs, some efforts have been made to achieve cost-effective LLM-based log parsers. LILAC\cite{jiang2024lilac} enhances the efficiency of LLM-based log parsing by incorporating an adaptive parsing
cache that stores log templates, Logbatcher\cite{xiao2024free} and LUNAR\cite{lunar} partition logs and only queries LLM once for logs belonging to the same cluster. 
\textit{InferLog} identifies the performance bottleneck in LLM inference under concurrent requests and optimized by ICL-oriented prefix caching, orthogonal to existing techniques for reducing the number of queries.

\textbf{LLM Inference Optimization.}
Existing inference engines, such as vLLM\cite{vllm} and SGLang\cite{sglang}, integrate numerous advanced techniques such as iteration-level scheduling\cite{orca}, PagedAttention\cite{vllm}, chunked prefill\cite{chunk} and speculative decoding\cite{speculative}. 
Reusing KV cache across different requests have been applied to recent works. 
Prompt Cache\cite{promptcache} allows flexible reuse of the same tokens at different positions.
Liu et al.\cite{databasekvcache} proposed
column reordering and row sorting to improve the prefix KV cache hit rate in relational analytical workload. RAGCache\cite{jin2024ragcache} improved the cache efficiency for RAG system. \textit{InferLog} optimizes prefix KV cache reusing for ICL paradigm and enhances performance based on configuration tuning.

\vspace{-0.2cm}
\section{Conclusion}
In this work, we propose \textit{InferLog}, a novel framework of optimizing LLM inference for online log parsing of software systems. Based on the characteristics of LLM-based log parsing tasks and LLM inference performance, \textit{InferLog} employs PAIR that consisted of prefix-based matching, modifying and reordering to refine ICL examples, thereby enhancing prefix KV cache hit rates. \textit{InferLog} further combines attention-augmented meta-learning with SMBO for rapid, tailored configuration optimization. Experiments confirm its effectiveness in enhancing inference efficiency while preserving log parsing quality.

\textbf{Artifact Availability}. The
source code and data is available at \url{https://github.com/wiluen/InferLog}.


\section{Acknowledgments}
The authors would like to thank the reviewers for their valuable comments and suggestions. This work was supported in part by National Key Research and Development Program of China (Grant Number: 2024YFB4505904), the National Natural Science Foundation of China under Grant 62272495 and the Guangdong Basic and Applied Basic Research Foundation under Grant 2023B1515020054. The corresponding author is Pengfei Chen.

\bibliographystyle{ACM-Reference-Format}
\bibliography{refs}

\end{document}